\documentclass[prd,eqsecnum,showpacs,aps,nofootinbib]{revtex4}
\usepackage{latexsym}
\usepackage{graphicx}
\newcommand{\del}{\partial}

\newcommand{\x}{{\mathbf{x}}}
\newcommand{\y}{{\mathbf{y}}}

\newcommand{\f}{\frac}
\newcommand{\T}{\tilde}

\newcommand{\bb}{\bibitem}
\newcommand{\BF}{\begin{figure}\begin{center}}
\newcommand{\EF}{\end{center}\end{figure}}
\newcommand{\BE}{\begin{equation}}
\newcommand{\EE}{\end{equation}}
\newcommand{\BEA}{\begin{eqnarray}}
\newcommand{\EEA}{\end{eqnarray}}
\newcommand{\ti}{\textit}

\begin{document}
\title{Harmonic Inpainting of the Cosmic Microwave Background 
Sky:\\Formulation and Error Estimate}

\author{Kaiki Taro Inoue}
\affiliation{Department of Science and Engineering, Kinki University,
Higashi-Osaka, Japan}
\author{Paolo Cabella}
\affiliation{Universita` 'La sapienza', Universita` 'Tor Vergata', Rome, Italy}
\author{Eiichiro Komatsu}
\affiliation{Department of Astronomy, The University of Texas at Austin, Texas, USA}

\date{\today}

\begin{abstract}
We develop a new interpolation scheme, based on harmonic inpainting,
for reconstructing the cosmic microwave background CMB temperature 
data within the Galaxy mask from the data outside the mask. We find that, 
for scale-invariant isotropic random Gaussian 
fluctuations, the developed algorithm reduces the errors in the
 reconstructed map for the odd-parity modes significantly
for azimuthally symmetric masks with constant galactic latitudes.
For a more realistic Galaxy mask, we find a modest improvement in
the even parity modes as well. 
\end{abstract}
\maketitle
\section{Introduction}

After the first release of the cosmic microwave background 
(CMB) anisotropy data from the Wilkinson Microwave Anisotropy Probe
(WMAP) in 2003, various types of ``anomalies'' in the CMB
temperature anisotropy  
on large angular scales have been
reported\cite{teg1,oli1,vie1,cru1,eri1,han1,chiang1,copi1,park1,schwarz1}.  
The origin of these anomalies -- whether they are cosmological,
statistical fluke, or something else -- is
unknown\cite{gordon1,jaffe1,tomita1,tomita2,moffat1,campanelli1, is1,is2}.

Recent analyses based on either a Bayesian or frequentist approach
show that the statistical significance level of 
some of the large-angle anomalies is
sensitive to the treatment of the Galactic sky cut\cite{kate1,naselsky1}. 
The lack of robustness mainly comes from the fact that we cannot use the
full sky information, but can only use a part of the sky outside the
Galaxy mask. There is a way to construct a full sky map that minimizes
the foreground contamination, e.g., the Internal 
Linear Combination (ILC) technique\cite{bennett/etal:2003,hinshaw/etal:2007};
however, we cannot rule out a potential residual foreground
contamination in the ILC map, especially in the region that is very
close to the Galactic plane, due to a limited number of sky maps at 
different frequencies. 

An alternative approach is to {\it reconstruct} the temperature data
within the Galaxy mask from the information available outside the mask.
So far, various types of methods such as
the direct inversion, the Wiener filtering,
and the power equalization filtering have been proposed for
reconstructing the CMB anisotropy on the cut sky\cite{bielewicz1}.
The Gibbs sampling \cite{gibbs1} provides a way to compute an ensemble  
of the Wiener filtered maps with a statistical sample of power spectra  
fit to a given input sky map, which can be used 
to estimate the best-fitting Wiener filtered map
and the corresponding uncertainties; see \cite{gibbs2} and references  
therein. All these methods are based upon a linear transformation 
of the expansion coefficients of spherical harmonics on the cut sky,
$a_{lm}^{cut}$, into the full sky coefficients, $a_{lm}^{full}$, which
are suitable for nearly all-sky data with a small sky cut. 
However, it has
been shown that the methods do not work properly for maps with a large
sky cut\cite{tegmark97, efstathiou1}.

In order to reconstruct the data within the Galaxy mask, one needs to
assume a prior on the properties of the data.
For instance, one may require the temperature anisotropies and their  
absolute values of the gradient to have a Gaussian distribution. Then, one can 
find an optimal solution for reconstructing the data
within the masked region from available information outside the
mask. Such an operation  
is called an ``inpainting'', which has been used
by skilled museum or art workers for restoring 
damaged photographs, films, and paintings. In recent years,  
various automatic inpainting algorithms based on 
partial differential equations or the variational principle have been proposed 
and used for automatic image restoration and removal of
occlusions\cite{bertalmio1, chan1, bertalmio2}.  
 
In this paper, we formulate an algorithm based on
harmonic inpainting for reconstructing smooth CMB data within
the Galaxy mask. In Sec. II, we develop a numerical 
scheme for implementing harmonic inpainting on a unit sphere
based on the boundary element method.
In Sec. III, we use Monte Carlo simulations to 
estimate the errors of the inpainted signal for
azimuthally symmetric sky cuts as well as for a realistic non-symmetric cut.
In Sec. IV, we summarize our results.

\section{Harmonic inpainting }

Let $\x=\x(\theta ,\phi)$ be a unit
pointing vector on the sky towards a given direction, $(\theta,\phi)$,
and $u_0(\x)$ be an observed temperature fluctuation outside the
Galaxy mask,  $\x\in \bar{D}$, where $D$ denotes the region within the
mask (an open connected set) and $\bar{D}$ outside the mask
(the compliment of $D$).
We assume that the boundary of $D$, 
$\del D$, is smooth. We wish to reconstruct 
a smooth temperature fluctuation on the full sky, or a 
best inpainting $u(\x)$, from the data outside the mask, $u_0(\x)$.
 
We reconstruct the full sky map, $u(\x)$, by locally minimizing the 
following ``energy'',
\BE
{\cal{H}}[u]=\int_{S^2} \left[\lambda(u-u_0)^2+|\nabla u|^2\right]
\sqrt{g }d v_{\x}, 
\label{eq:H}
\EE
where $\lambda$ is a positive
constant, $g_{ij}$ is the Riemannian metric tensor of a sphere, and
$dv$ is the infinitesimal Euclidean volume. 
The first term and the second 
term in the r.h.s of Eq. (\ref{eq:H}) represent 
faithfulness to the imperfect data and the 
regularization penalty, respectively. 
The constant, $\lambda$, controls a tradeoff
between faithfulness to the imperfect data and 
smoothness of the fluctuations.
This procedure is optimal when the absolute values of 
the gradient, $|\nabla u|$, and the difference between $u$ and $u_0$ 
are Gaussian distributed\footnote{
This is a reasonable requirement for Gaussian random 
fluctuations as each component of the 
gradient is also a Gaussian random variable.
}. 
The Euler-Lagrange equation of (\ref{eq:H})
is given by
\BE
-\Delta u+\lambda (u-u_0)=0; ~~~ \lambda>0~~~~\textit{for}~~~\x \notin D, ~~~ 
\lambda=0~~~ \textit{for}~~~ \x \in D. \label{eq:EL}
\EE
In order to solve Eq. (\ref{eq:EL}), 
we also need a solution for the derivative of $u$ with respect to 
the unit vector $n$ normal to the boundary of the mask, 
\BE
q\equiv\frac{\del u}{\del n}=\left(\frac{\del u}{\del x^i}\right) n^i, 
\EE
where $\x\in \del D$ for which $u$ and $q$ are continuous. 

The minimizer of Eq.(\ref{eq:H}) is called the harmonic
inpainting denoizing of $u_0$. 
Expanding a temperature fluctuation in real spherical harmonics as
$u=\sum a_{lm} Y_{lm}$ and 
$u_0=\sum b_{lm} Y_{lm}$, 
the solution of Eq.(\ref{eq:EL}) for the region outside the mask,
$\x\in \bar{D}$, is given by  
\BE
a^{out}_{lm}(\lambda)=\f{\lambda}{l(l+1)+\lambda} b_{lm}, \label{eq:3}
\EE
where $a^{out}_{lm}$'s are real expansion coefficients for the solution
outside the mask.  Eq.(\ref{eq:3}) implies a suppression of 
modes with $l \gg \sqrt{\lambda}$. In the limit of $\lambda \gg 1$
\footnote{This corresponds to the case where the Gaussian 
noise of the fluctuation $u-u_0$ outside the mask is negligible in
comparison with the expected 
absolute values of gradients $|\nabla u|$. For instance, from
the signal-to-noise ratio, one can estmate that 
$\lambda \gtrsim 10^4$ for the WMAP data at $l<15$.}, 
the minimizer of Eq.(\ref{eq:EL}) is called the 
real harmonic inpainting.

The fluctuation, $u(\x)$,
 satisfies the following boundary integral equation (see
 \cite{Inoue2001} for further detail), 
\BE
\f{u(\x)}{2} +\int_{\del D} G(\x,\y) q(\y) \sqrt{g} d\y
  - \int_{\del
D} H(\x,\y) u(\y) \sqrt{g}d\y=0,~~~~ \x\in \del D,
\label{eq:BEM}
\EE
where
$G$ is the Green's function of Laplacian
$\Delta$, and $H$ is the normal derivative of $G$,
$H\equiv \del G /\del n$. 
By discretizing the boundary, $\del D$, into $2 N$ elements, $\Gamma_j$, 
and approximating $u$ on $\del D$ 
by some low-order polynomials, Eq.(\ref{eq:BEM}) yields 
\begin{equation}
[H]\{u\}=[G]\{q\},~~~~q\equiv \f{\del u}{\del n},\label{eq:HG}
\end{equation}
where $\{u\}$ and $\{q\}$ are $N$-dimensional vectors consisting of 
the boundary values of $u_i$ and their normal derivatives, $q_i$,
and $[H]$ and $[G]$ are 
$N\times N$-\,dimensional coefficient matrices which are
obtained from the integration of 
the fundamental solution, $G(\x,\y)$, and its normal derivatives,
$H(\x,\y)$. For constant elements
\begin{equation}
u(\x_j)=u^j=Const.\,,~~~~q(\x_j)=q^j=Const.\,~~~ 
\mathrm{on}~~ \Gamma_j,
\end{equation}  
the elements of the matrices are 
\begin{eqnarray}
H_{ij}&=&\left\{
\begin{array}{@{\,}ll}
\T{H}_{ij} & \mbox{$i \neq j$}\\
\T{H}_{ij}-\f{1}{2} & \mbox{$i=j$}, 
\nonumber
\end{array}
\right.
\end{eqnarray}
where 
\begin{equation}
\T{H}_{ij}\equiv \int_{\Gamma_{j}}\f{\del G}{\del n}
(\x_i,\y_j) \,\sqrt{g}d \y,~~G_{ij}\equiv  \int_{\Gamma_{j}}
G(\x_i,\y_j) \,\sqrt{g}d \y. 
\end{equation}

In terms of the normal derivatives $\{q\}$ that are derived from 
Eq.(\ref{eq:HG}), the reconstructed function $u$ within the masked region,
$D$, is given by
\BE
u(\x)=-\int_{\del D} G(\x,\y) q(\y) \sqrt{g} d\y
  + \int_{\del
D} H(\x,\y) u(\y) \sqrt{g}d\y, ~~~~\x \in D.
\EE

\section{Error estimation}
\subsection{Application to azimuthally symmetric masks}
In order to check the robustness of the algorithm
developed in section II, we carry out Monte-Carlo 
simulations of random Gaussian fluctuations on a unit sphere, $u_g$.
We generate Monte Carlo realizations from a scale invariant 
power spectrum ($n=1$), 
\BE
C_l=\langle |a_{lm}|^2\rangle\propto \f{1}{l(l+1)},
\EE
which describes the large-angle CMB temperature
fluctuations in the Einstein-de Sitter (EdS) 
universe. 

We first consider azimuthally symmetric masks with 
constant galactic latitudes, i.e., we mask region 
below given galactic latitudes,  $|b|<\theta_0$.
For isotropic
random fluctuations, modes with $l\gtrsim \pi/\theta_0$ are not affected
by the azimuthal cut very much (except for the overall
normalization). Therefore, in the following we shall use 
only multipoles with $l<l_c$, where $l_c = \pi/\theta_0$.
From Eq.(\ref{eq:3}), one can see that this choice of 
cutoff in harmonic space is roughly equivalent to 
choosing the control parameter, $\lambda$, such that
$\lambda=(\pi/\theta_0)^2=l_c^2$. The number of elements
on the boundary necessary for numerically solving the integral equation 
(\ref{eq:BEM}) depends on the size of the mask.
For instance, we discretize the boundary
of the mask into 80 linear elements for $\theta_0=20^\circ$.  

To measure the difference 
between the reconstructed fluctuation, $u(\x)$, and the original one, 
$u_g(\x)$, we  
use the $L^2$-norm defined as 
\begin{equation}
(a[u])^2\equiv \int_{S^2} (u[\theta,\phi])^2 d\Omega
=\sum_{lm}(a_{lm}^{\small{\textrm{HI}}})^2,
\end{equation}
where $d\Omega$ is the infinitesimal area element for a surface of unit sphere,
and $a_{lm}^{\rm HI}$ is the spherical harmonic coefficients of the
reconstructed map. 
To reduce the sample variance, we generate
1000 realizations, each of which is masked by the azimuthally symmetric
mask and is reconstructed by the developed algorithm.
We then calculate the mean relative errors defined as
\begin{equation}
\Bigl (\f{\Delta a}{a}\Bigr)^2\equiv \Biggl \langle  
\f{\int_{S^2} (u_g[\theta,\phi]-u[\theta,\phi])^2 d\Omega}
{\int_{S^2} (u_g[\theta,\phi])^2 d\Omega}  \Biggr  \rangle, 
\end{equation}
where $u_g(\x)$ and $u(\x)$ are the underlying Gaussian fluctuation
and the inpainted (reconstructed) fluctuation, respectively. For comparison, 
we also compute the relative errors
for the fluctuations with the ``naive'' ansatz in which $u=0$ 
within the masked region and $u=u_0$ outside the mask.

\begin{figure}[t]
\includegraphics[width=14cm]{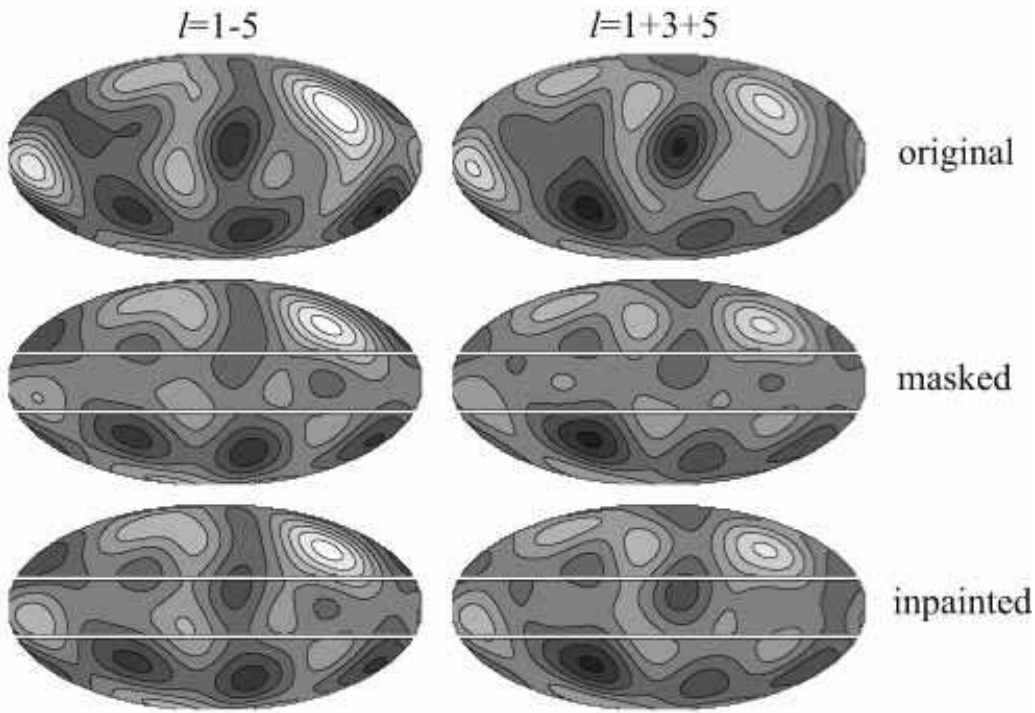}
\caption{Contour maps of one Gaussian 
realization (top), those with an azimuthally symmetric mask 
$|b|<20^\circ$ (middle),  
and the inpainted maps (bottom)
in Mollweide projections. The left figures show co-added
multipoles (from $l=1$ to $l=5$) whereas the right figures show 
co-added odd multipoles ($l=1,3,5$). The boundaries of the mask
are shown in white lines. }
\label{fig:realization1}
\end{figure}

As one can see in Fig.{\ref{fig:realization1}}, 
some of the missing features within the masked region 
are reconstructed by the inpainting. Improvements are
more conspicuous for odd multipoles. 

In Table \ref{tab:1}, we compare the mean relative errors, $(\Delta
a/a)^2$, at $l=1$, 2, 3, 4, and 5, for two azimuthally symmetric masks,
$|b|<20^\circ$ and $30^\circ$.
Compared with the naive cases, our developed algorithm based on 
harmonic inpainting significantly improves 
reconstruction of odd multipoles. 
However, for azimuthally symmetric masks, we find that the inpainted
even multipoles are identical with those in the naive case.

This can be explained as follows. The 
Green's function on a unit sphere is given by
$G(\x,\y)=-Q_0(\x\cdot \y)/2 \pi$, where $2Q_0(\x \cdot 
\y)=\ln [(1+\x \cdot \y)/(1-\x\cdot \y)]$; thus, 
we have $G(\x,\y)=G(-\x,-\y)$. In other words,
the Green's function is invariant under the spatial 
inversion. Therefore, if the boundary of the masked region 
is invariant under the spatial inversion, the order of the $2N$-dimensional 
element matrix $[G]$ is reduced to $N$, 
which kills the degree of freedom for even-parity modes. 

\begin{figure}[t]
\includegraphics[width=9cm]{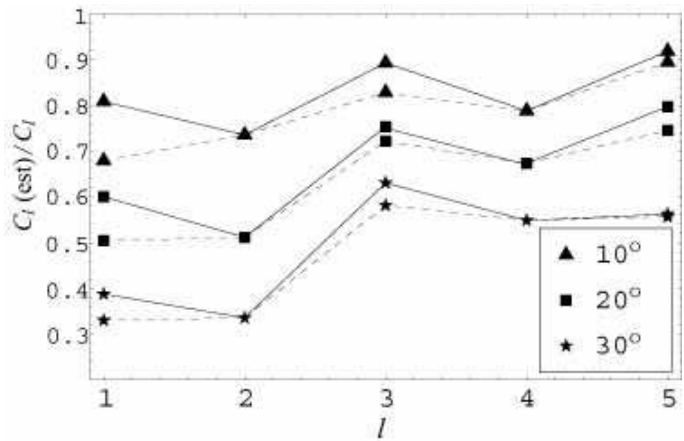}
\caption{The estimated (pseudo) power spectrum, $C_l({\rm est})=\sum
 |a_{lm}|^2/(2l+1)$, for azimuthally symmetric masks with
 $|b|<\theta_0=10^\circ$, $20^\circ$, and $30^\circ$, divided by the
 input $C_l\propto (l(l+1))^{-1}$. The estimated $C_l({\rm est})$ have
 been averaged over 1000 Monte Carlo realizations, and  plotted for the
 inpainted (reconstructed) fluctuations (solid lines) and for the naive
 ones (dashed lines).   
}
\label{fig:bias}
\end{figure}

The sky cut biases the estimation of the angular power spectrum when it
is computed naively as $C_l(\textrm{est})=\sum_m
(a_{lm}^{\small{\textrm{HI}}})^2/(2l+1)$, which is often called the
``pseudo $C_l$'' \cite{pseudo-cl}.
The effect of mask needs to be
deconvolved using, e.g., the MASTER method \cite{master}.
Here, we do not use the MASTER method, but study how much the bias is
reduced when we use the naive $C_l$ estimation and the harmonic
inpainting.
In Figure \ref{fig:bias} we show the ratio of the estimated $C_l$
divided by the input $C_l$, averaged over 1000 Monte Carlo simulations,
for three azimuthally symmetric masks, $|b|<10^\circ$, $20^\circ$, and
$30^\circ$. We find modest improvements at odd multipoles, whereas no
improvements are found at even multipoles as they are identical to
those in the naive case.

\begin{table}[t]
\begin{center}
\begin{tabular}{|c|c|c|c|c|c|}   \hline
\multicolumn{6}{|c|}{$|b|<20^\circ $}\\
\hline
\multicolumn{1}{|c|}{$l$ } &
\multicolumn{1}{|c|}{1 } &
\multicolumn{1}{|c|}{2} &
\multicolumn{1}{|c|}{3} &
\multicolumn{1}{|c|}{4}&
\multicolumn{1}{|c|}{5}

 \\  \hline
 naive   &  0.23 & 0.21 & 0.55 & 0.39 & 0.55 \\
 \hline
 inpainted   & 0.13 &0.21 & 0.31 & 0.39 &0.32 
       \\ \hline 
\multicolumn{6}{|c|}{$|b|<30^\circ $}\\
\hline
\multicolumn{1}{|c|}{$l$ } &
\multicolumn{1}{|c|}{1 } &
\multicolumn{1}{|c|}{2} &
\multicolumn{1}{|c|}{3} &
\multicolumn{1}{|c|}{4}&
\multicolumn{1}{|c|}{5}

 \\  \hline
 naive   &  0.39 & 0.37 & 0.76 & 0.58 & 0.62 \\
 \hline
 inpainted   & 0.27 &0.37 & 0.50 & 0.58 &0.48 
       \\ \hline     
\end{tabular}\caption{Mean relative errors, $(\Delta a/a)^2$, of
 multipoles for azimuthally symmetric masks with constant galactic latitudes} 
\label{tab:1}
\end{center} 
\end{table}

\subsection{Application to a realistic Galaxy mask}
As we have seen, the developed algorithm based on 
harmonic inpainting fails to reconstruct
even multipoles when the Galaxy mask is azimuthally
symmetric. However, the shape of the Galactic foreground emission region
is not perfectly azimuthally symmetric, and thus the 
corresponding Galaxy mask does not respect parity conservation. 
Therefore, we expect an improvement for even multipoles as well relative
to the naive estimations, once 
realistic Galaxy masks are used. 

To confirm this expectation, 
we have carried out Monte-Carlo simulations with a realistic Galaxy
mask.
In this study we also use a realistic sky signal as well: 
we have generated 1000 realizations of temperature maps 
from a  $\Lambda$CDM model
with cosmological parameters given by 
$(n,\Omega_m,\Omega_\Lambda)=(1,0.24,0.76)$.
As for the Galaxy mask, we use the WMAP's {\it Kp0}
mask\cite{bennett/etal:2003}. 

Since the shape of the {\it Kp0} mask is fairly irregular at the scale of pixels,
the definition of the border is not so trivial like the azimuthally symmetric
mask in Fig. 1. For instance, some observed pixels close to 
the mask are surrounded by two or more pixels, 
some others are completely surrounded by masked
pixels, making the choice of the border arbitrary and unstable. To avoid
this we have built a smoothed version of the {\it Kp0} border.
Our procedure consists
of going (for a fixed $\phi$ sampled with a constant interval) 
along the coordinate $\theta$ from Galactic north to south 
until we find a masked pixel belonging to the ${\it Kp0}$ mask. 
Once we have found this pixel, we would go back by a step of 10
arcminutes 
and choose that pixel living on the border. The final set of
($\theta_i,\phi_i$) defines our border. 
We found that this procedure regularizes the {\it Kp0} border.

\begin{figure}[t]
\includegraphics[width=8cm]{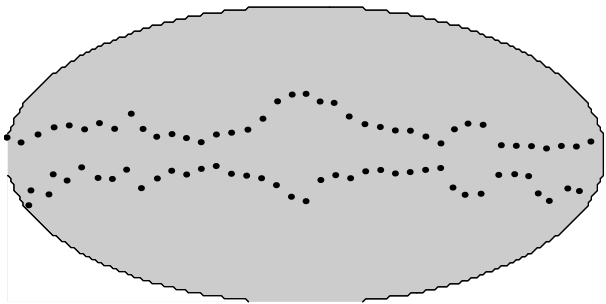}
\caption{Meshes on the boundary of the {\it Kp0} Galaxy mask, smoothed
on scale $\sim 10$ arcmin.}
\label{fig:Kp0-meshes}
\end{figure}

We choose the cutoff angular scale to be 
$l_c=15$, which corresponds to the mean  
angular size of the mask in the polar direction.
The boundary of the {\it Kp0} mask are discretized
into 80 linear elements ($\theta_i,\phi_i$) 
with a constant interval $2\pi/40$ (Fig.\ref{fig:Kp0-meshes}). 

\begin{figure}[t]
\includegraphics[width=17cm]{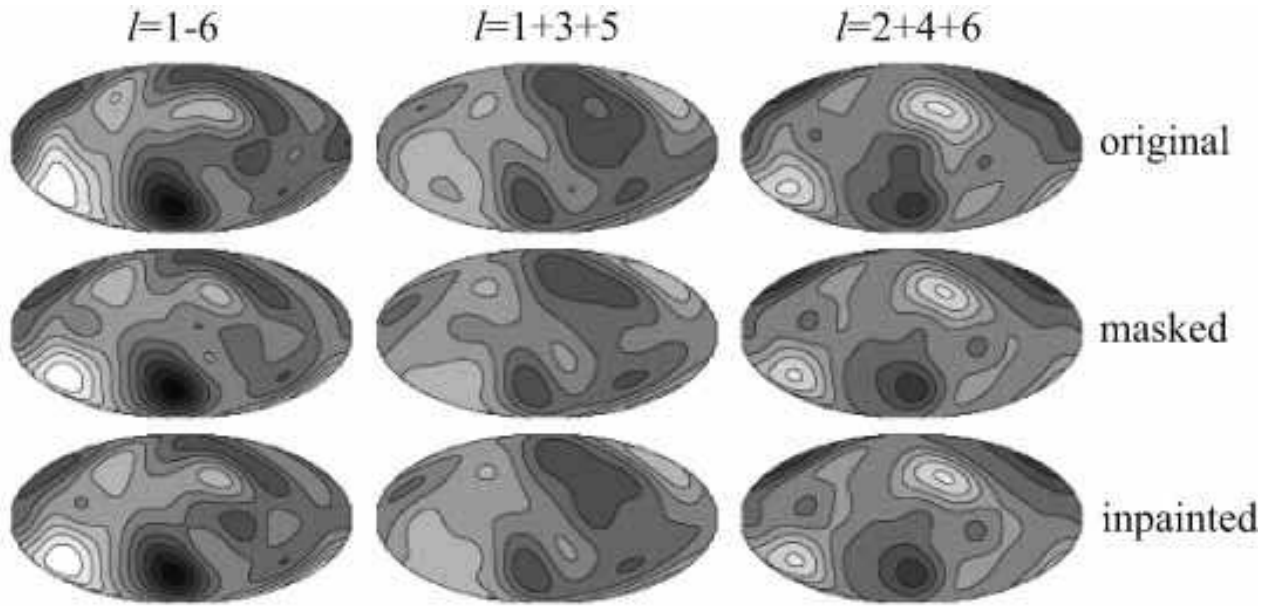}
\caption{Contour maps of one Gaussian 
realization (top), those with a \ti{Kp0} mask (middle) 
and the inpainted maps (bottom)
in Mollweide projections with galactic coordinates.
The left figures show co-added
multipoles (from $l=1$ to $l=6$) whereas the middle and right 
figures show co-added odd multipoles ($l=1,3,5$) and 
co-added even multipoles ($l=2,4,6$), respectively. }
\label{fig:realization2}
\end{figure}

As in the case of the azimuthally 
symmetric mask, some of the missing features within the {\it Kp0} mask
are reconstructed by the inpainting. The improvements are
more conspicuous for odd multipoles than even multipoles 
(Fig.\ref{fig:realization2}).

We show the mean relative errors for the naive estimation and those 
for the inpainted map with the {\it Kp0} mask in Table II. 
We find that our algorithm improves reconstruction of 
both the odd and even multipoles, but the improvement for the even
multipoles is modest (3 to 10\%). 


In order to find the polar-angle dependence, we compute
the mean relative errors for each ($l,m$) mode,
defined by
\begin{equation}
\Bigl (\f{\Delta a}{a}\Bigr)^2_{(l,m)}\equiv \Biggl \langle  
\f{(a_{lm}^{\textrm{\small{HI}}}- a_{lm})^2}
{(a_{lm}^{\textrm{\small{HI}}})^2} \Biggr  \rangle, 
\end{equation}
where $a_{lm}^{\textrm{\small{HI}}}$ and $a_{lm}$
are real expansion coefficients for the inpainted map
and the original map, respectively. 

\begin{figure}[t]
\includegraphics[width=12cm]{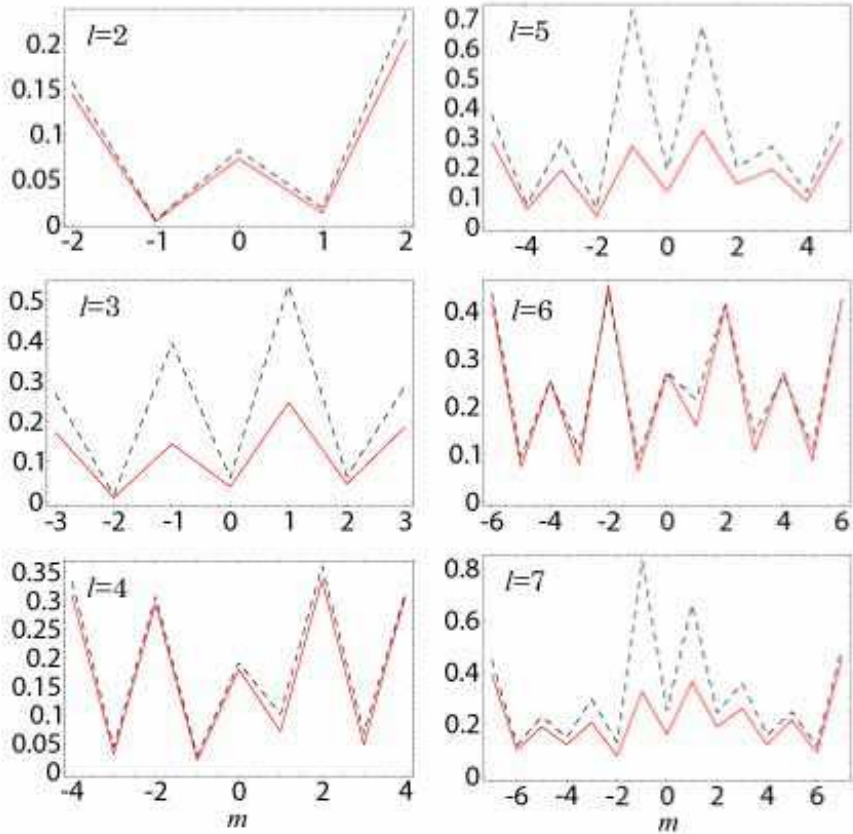}
\caption{Mean relative errors $(\Delta a/a)^2_{(l,m)}$ 
for the {\it Kp0} Galaxy mask. The solid lines and the dashed lines
 show the inpainted and naive cases, respectively.  }
\label{fig:Kp0-errors-lm}
\end{figure}

\begin{figure}[t]
\includegraphics[width=9cm]{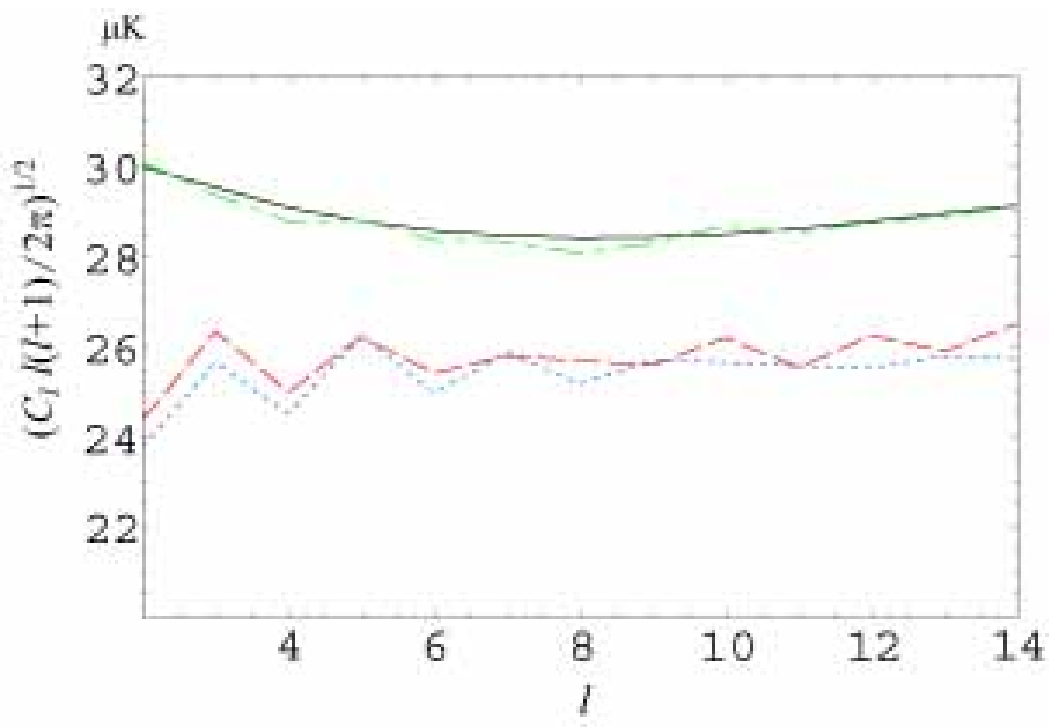}
\caption{
The estimated (pseudo) power spectrum, $C_l({\rm est})=\sum
 |a_{lm}|^2/(2l+1)$, for the {\it Kp0} Galaxy mask, averaged over 1000
 Monte Carlo realizations.
 The dashed and dotted lines show $C_l({\rm est})$ from the inpainted
 map and the map with the {\it Kp0} Galaxy mask, respectively.  
The solid and the dash-dotted lines show the input $C_l$ 
and the full-sky $C_l({\rm est})$ estimated from  1000 realizations, 
respectively.
}
\label{fig:bias-kp0}
\end{figure}

We find that significant improvements  
come mainly from odd-parity modes with $m=\pm1$
(Fig. \ref{fig:Kp0-errors-lm}).
 Lower multipoles
are improved more by inpainting than higher ones. 
We also find that the biases in the pseudo $C_l$ are reduced for both 
the odd and even multipoles (Fig. \ref{fig:bias-kp0}).  

\begin{table}[t]
\begin{center}
\begin{tabular}{|c|c|c|c|c|c|c|c|c|c|}   \hline
\multicolumn{1}{|c|}{$l$ } &
\multicolumn{1}{|c|}{1 } &
\multicolumn{1}{|c|}{2} &
\multicolumn{1}{|c|}{3} &
\multicolumn{1}{|c|}{4}&
\multicolumn{1}{|c|}{5}&
\multicolumn{1}{|c|}{6} &
\multicolumn{1}{|c|}{7} &
\multicolumn{1}{|c|}{8} &
\multicolumn{1}{|c|}{9}

 \\  \hline
 naive   &  0.096 & 0.11 & 0.28 & 0.23 & 0.35 & 0.28 & 0.35 & 0.30 & 0.33 \\
 \hline
 inpainted   & 0.053 &0.10 & 0.15 & 0.21 & 0.21 & 0.27 & 0.24 & 0.29 & 0.25 
       \\ \hline 
\end{tabular}\caption{Mean relative errors $(\Delta a/a)^2$ of
multipoles for the {\it Kp0} Galaxy mask.} 
\label{tab:Kp0-errors}
\end{center} 
\end{table}

\section{Summary}
We have shown that our new algorithm based on 
harmonic inpainting offers a way to 
reconstruct the missing CMB temperature anisotropy data within 
the Galaxy mask, provided that the primordial fluctuation 
is Gaussian with a scale-invariant ($n=1$) spectrum. 
The method reduces the errors for 
multipoles with an odd-parity (odd $l$) significantly.
While the method fails to reconstruct 
multipoles with an even parity (even $l$) for azimuthally symmetric
masks, it can reduce the errors for multipoles with an even parity 
for a more realistic,
non-azimuthally symmetric mask, such as the WMAP's {\it Kp0} mask.
As we have shown, our 
method is useful for studying the dependence of the CMB 
at low multipoles given by the azimutal parameter m of the harmonic
coefficients, which is relevant to confirm or confute
some of the anomalies at low multipoles found in the WMAP data.

Our new algorithm does not require any specific form 
of the power spectrum as a prior for reconstructing the missing
data as long as the fluctuations obey a Gaussian distribution.  
Even when fluctuations are not Gaussian, 
our algorithm is expected to work as long as the fluctuations are
sufficiently smooth and the gradients approximately 
obey a Gaussian distribution.  
Therefore, our method may be used to reconstruct other 
observational data that is sufficiently smooth 
such as radio or infrared maps on a cut sky (a similar
technique that can incorporate texture as well as the smooth part
is studied in \cite{Abrial07}).

Our analysis implies that low multipoles with an even parity
are more likely to suffer from the effect of sky cut compared
with those with an odd parity. In other words, even 
parity modes on large angular scales cannot be precisely 
reconstructed without assuming specific forms of 
the angular power. Therefore, the statistical
significance of the anomalous features involving the CMB 
quadrupole may be much lower than claimed, as recent studies 
suggest\cite{kate1, naselsky1}. 

The feature of the inpainted 5 year WMAP data will be explored 
in the forthcoming paper (Inoue \& Cabella 2008).

\end{document}